\documentclass[twocolumn]{aastex63}

\usepackage{amsmath}
\usepackage{graphicx}
\usepackage{txfonts}

\received{\today}
\revised{}
\accepted{}
\submitjournal{AJ}

\shorttitle{Radiative scale-height}
\shortauthors{Montesinos et al.}
\graphicspath{{./}{figures/}}

\begin{document}

\title{Radiative scale-height and shadows in protoplanetary disks}

\correspondingauthor{Matías Montesinos}
\email{matias.montesinos@uv.cl}

\author{Matías Montesinos}
\affiliation{Instituto de F\'isica y Astronom\'ia, Universidad de Valpara\'iso, Chile}
\affiliation{Chinese Academy of Sciences South America Center for Astronomy, National Astronomical Observatories, CAS, Beijing 100012, China}
\affiliation{N\'ucleo Milenio de Formaci\'on Planetaria (NPF), Chile}

\author{Nicolás Cuello}
\affiliation{Univ. Grenoble Alpes, CNRS, IPAG, F-38000 Grenoble, France}
\affiliation{Instituto de Astrof\'isica, Pontificia Universidad Cat\'olica de Chile, Santiago, Chile}
\affiliation{N\'ucleo Milenio de Formaci\'on Planetaria (NPF), Chile}

\author{Johan Olofsson}
\affiliation{Instituto de F\'isica y Astronom\'ia, Universidad de Valpara\'iso, Chile}
\affiliation{N\'ucleo Milenio de Formaci\'on Planetaria (NPF), Chile}

\author{Jorge Cuadra}
\affiliation{Departamento de Ciencias, Facultad de Artes Liberales, Universidad Adolfo Ib\'a\~nez, Av.\ Padre Hurtado 750, Vi\~na del Mar, Chile}
\affiliation{N\'ucleo Milenio de Formaci\'on Planetaria (NPF), Chile}

\author{Amelia Bayo}
\affiliation{Instituto de F\'isica y Astronom\'ia, Universidad de Valpara\'iso, Chile}
\affiliation{N\'ucleo Milenio de Formaci\'on Planetaria (NPF), Chile}

\author{Gesa H.-M. Bertrang}
\affiliation{Max Planck Institute for Astronomy, Königstuhl 17, 69117 Heidelberg, Germany}

\author{Clément Perrot}
\affiliation{Instituto de F\'isica y Astronom\'ia, Universidad de Valpara\'iso, Chile}
\affiliation{N\'ucleo Milenio de Formaci\'on Planetaria (NPF), Chile}
\affiliation{LESIA, Observatoire de Paris, Universit\'e PSL, CNRS, Sorbonne Universit\'e, Univ. Paris Diderot, Sorbonne Paris Cit\'e, 5 place Jules Janssen, 92195 Meudon, France}



\begin{abstract}

   Planets form in young circumstellar disks called 
protoplanetary disks. However, it is still difficult to catch planet formation in-situ. Nevertheless, from recent ALMA/SPHERE data, encouraging evidence of the direct and indirect presence of embedded planets has been identified in disks around young stars: co-moving point sources, gravitational perturbations, rings, cavities, and emission dips or shadows cast on disks.

  The interpretation of these observations needs a robust physical framework to deduce the complex disk geometry. In particular, protoplanetary disk models usually assume the gas pressure scale-height given by the ratio of the sound speed over the azimuthal velocity $H/r = c_{s\rm }/v_{\rm k}$. By doing so, \textit{radiative} pressure fields are often ignored, which could lead to a misinterpretation of the real vertical structure of such disks.
 
   We follow the evolution of a gaseous disk with an embedded Jupiter mass planet through hydrodynamical simulations, computing the disk scale-height including radiative pressure, which was derived from a generalization of the stellar atmosphere theory.  We focus on the vertical impact of the radiative pressure in the vicinity of circumplanetary disks, where temperatures can reach $\gtrsim 1000$ K for an accreting planet, and radiative forces can overcome gravitational forces from the planet.
 
   The radiation-pressure effects create a vertical optically thick column of gas and dust at the proto-planet location, casting a shadow in scattered light. This mechanism could explain the peculiar illumination patterns observed in some disks around young stars such as HD~169142 where a moving shadow has been detected, or the extremely high aspect-ratio $H/r \sim 0.2$ observed in systems like AB Aur and CT Cha.

\end{abstract}


\section{Introduction}

During the last years, high angular resolution imaging  of disks around young stars has revealed different patterns in the gas and dust structures. These structures are usually observed in gas-rich disks with an inner cavity around low and intermediate-mass stars, classified as transition disks. Spatially resolved observations reveal astonishing images of circumstellar disks. For instance, in the disk around HL Tau ALMA unveiled for the first time a series of gaps at millimeter wavelengths \citep{ALMA2015}. Several remarkable other examples from high angular resolution images (at near-infrared and sub-mm wavelengths) show multiple gaps, spirals, and rings. It is thought that embedded planets cause most of these features (e.g., \citealt{Long+2018ApJ,Keppler+2018}). A complete review of disk substructures' observations at high angular resolution can be found in the DSHARP project paper series \citep{Andrews+2018, Huang+2018ApJ, Huang+2018ApJb}.
 
Besides these astonishing gas/dust structures, intriguing illumination features --- both in scattered light and thermal (sub)millimeter emission --- have been observed. For instance, \cite{Avenhaus+2014} reported two intensity nulls seen in the circumbinary disk of HD~142527. These were later identified as shadows cast by a tilted inner circumprimary disk \citep{Marino+2015, Casassus+2015ApJ}. Moreover, multiple shadow features had been discovered through high-contrast polarimetric differential imaging (PDI) with VLT/SPHERE in the outer disk of HD~135344B \citep{Stolker+2016} and the disk around TW Hya \citep{Debes+2017}.

Interestingly, \cite{Quanz+2013} reported VLT/NACO observations of the disk around the isolated Herbig~Ae/Be star HD~169142 showing a narrow emission dip located at $\sim 80^\circ$. More recently, \cite{Bertrang+2018, Bertrang+2020} presented VLT/SPHERE/ZIMPOL observations of the same system at higher spatial resolution. The latter reveals an inner ring at 24~au from the star and a narrow emission dip in surface brightness located now at $\sim50^{\circ}$. Remarkably, the \textit{moving shadow} could be caused by a rotating optically thick bump possibly located at $\sim 12$ au from the star, as suggested by \cite{Bertrang+2018}. 
    
Most of these features seem to be related to planet formation processes, in which planet-disk interactions sculpt the disk surface and mid-plane. However, direct detection of planets embedded in disks remains elusive. To date, only one planetary-mass companion has been imaged coexisting with a protoplanetary disk \citep{Keppler+2018, Muller+2018}, corresponding to the discovery of a forming planet within the gap in PDS~70, while some planet candidates have been proposed through kinematic signatures \citep{Pinte_2019, Pinte+2020}. 

An important quantity that characterizes the circumstellar disk geometry is its pressure scale-height, which is needed to explain several observations (especially the non-resolved ones). For instance, transitional disks exhibit an excess of NIR emission related to unusually high aspect-ratios ($H/r \sim 0.2$) at the inner rim, where a puffed-up wall is created by dust evaporation \citep{Natta+2001, Dullemond2001, Dullemond&Monnier2010, Olofsson+2013}. There is also some inferred difference in the aspect-ratio between the disks of groups I and II stars \citep{Meeus+2001}. Variations on the vertical scale-height in flat or only moderately flared disks are also supposed to be responsible for self-shadowing effects \citep{Garufi+2014, Stolker+2016}

The vertical disk extension is directly related to the fluid pressure field and gravity. In general, circumstellar gaseous disks are expected to be gas pressure-dominated. This condition is used to assume equilibrium between gravity and gas pressure, leading to the well-known relation for the aspect ratio $H/R = c_{\rm s} / v_{\phi}$, where $c_{\rm s}$ and $v_{\phi}$ are the sound speed and the azimuthal velocity of the gas, respectively. 

However, it is worth asking if this condition holds everywhere in the disk --- especially at high temperatures. In particular, upon which circumstances (if any) does the scale-height formula break down? In such a case, what kind of equation should be used instead? 

One important modification should be implemented in the vicinity of a planet, where its gravity dominates and must be taken into account \citep{Muller+2012}. Also, very high temperatures are expected to develop in some regions of protoplanetary disks. One of these hot regions arises at the inner boundaries of transitional disks where the dust sublimates, reaching temperatures in the range of $1000-1500$~K, which produces a puffed-up inner rim \citep{Natta+2001}. Also, highly luminous ($10^{-6} - 10^{-3} ~  \rm L_\odot$) forming planets are expected to form during early stages ($\sim 1 $ Myr) of planet formation \citep{Mordasini+2012}. From numerical simulations of disks with embedded luminous planets, \cite{Montesinos+2015} showed that the circumplanetary disk (CPD) region reaches temperatures of $\sim 1000$~K or more, with thermal emission peaking in L' in concordance with observations \citep{Spiegel+2012, Zhu+2015}. 

At high temperatures ($\geq 1000 $ K), the radiative pressure may becomes relevant, where processes similar to stellar atmospheres arise \citep{Mihalas+1984}. A simplified calculation of the scale-height, including radiative pressure, can be found in \cite{Montesinos2011}.  In that work, the authors also consider contributions from pressure fields related to viscous turbulence and the disk self-gravity, in the context of accretion disks around super-massive black holes.

In this paper, we adopt a generalization of the gray model for stellar atmospheres applied to disk geometry introduced by \cite{Hubeny1990} to include radiation pressure when computing the vertical structure of the disk. We also include the gravitational influence of an embedded planet in that calculation. Around hot accreting planets, radiative forces can dominate over the planet's gravity, creating an optically thick bump above the circumplanetary region. We link such bumps with self-shadowing effects through radiative transfer calculations. This could explain, for instance, the moving shadow observed in HD~169142 by associating it to the presence of an inner planet \citep{Bertrang+2018, Bertrang+2020}.

In Section \S{\ref{physics}}, we discuss the physical background of the atmosphere model to compute the disk scale-height. In \S{\ref{numerical}}, we run numerical hydro-simulations following the evolution of a protoplanetary disk with the adopted formulation. In \S{\ref{observations}}, we use our hydro-models to interpret observations by performing radiative transfer calculations on them. Finally, in \S{\ref{discussion}} we discuss our results, which are summarized in \S{\ref{conslusions}}.

\section{Scale-height}\label{physics}

The vertical structure of a disk can be computed under the assumption of hydro-static equilibrium between pressure forces and gravity. i.e., the vertical gradient pressure is balanced by the vertical tidal acceleration towards the mid-plane. Then, we have:
\begin{equation}\label{hidroeq}
\frac{{\rm d}P}{{\rm d}z} = -\int \rho \nabla \Psi dz ,
\end{equation}
where $z$ corresponds to the distance from the mid-plane, $P$ and $\rho$ are the total pressure and density, respectively. $\Psi$ is the gravitational potential given by contributions from the star, and the embedded planet, namely, $\Psi = \Psi_* + \Psi_{\rm planet}$.

\subsection{Scale-height near the planet}

Taking into account the gravitational potential from the star and the planet, the vertical hydro-static equation \ref{hidroeq} can be written in cylindrical coordinates as:

\begin{equation}\label{gradp}
    \frac{1}{\rho} \frac{\partial P}{\partial z} = - \frac{G M_*}{(r^2 + z^2)^{(3/2)}} - \frac{G M_{\rm p}}{(s^2 + z^2)^{(3/2)}},
\end{equation}
where $M_*$, and $M_p$ are the mass of the star and planet, respectively, and $s$ is the distance from the projected position of the planet ($z=0$ plane) to a disk element.

Assuming pure gas pressure given by $P_g = \rho c_s^2$, the scale-height $H$  in the absence of a planet is defined as: 

\begin{equation}\label{Hstd}
    H = \frac{c_{\rm s}}{\Omega},
\end{equation}
where $c_{\rm s}$ is the vertically isothermal sound speed, and $\Omega$ the angular velocity of the disk.

The embedded planet changes the disk structure, leading to a reduced thickness near it. We follow the procedure by \cite{Muller+2012} to compute the scale-height $H_{\rm planet}$, and the density $\rho_{\rm planet}$ of the disk in the vicinity of the planet. For a vertically isothermal disk, equation \ref{gradp} can be integrated:

\begin{equation}
    \rho_{\rm planet} = \rho_0 \exp{ \left\{-  \left(\frac{1}{2} \frac{z^2}{H^2} + \frac{|z|}{H_{\rm planet}}  \right)  \right\}},
\end{equation}
where the scale-height near the planet is given by,

\begin{equation}\label{Hplanet}
    H_{\rm planet} = \frac{4 s^2 H^2}{q r^3},
\end{equation}
where $q = M_p / M_*$, and $H$ defines the scale-height in the absence of the gravitational influence of the planet (Eq. \ref{Hstd}).

The planet influence is characterized by the transition distance \citep{Muller+2012}:

\begin{equation}
    s_t = \frac{1}{2} \left( \frac{r^3 q}{H}  \right)^{1/2}.
\end{equation}

When $s < s_t$, the planet dominates and therefore $H = H_{\rm planet}$ (Eq. \ref{Hplanet}), otherwise $H$ is given by equation \ref{Hstd}.

\subsection{Radiative scale-height}

In some cases, radiation pressure may be relevant, such as in circumplanetary disks around accreting planets where temperatures could reach high temperatures above 1000 K. We use an analytical model of the scale-height derived by \cite{Hubeny1990} to compute the vertical structure, obtained from a generalization of the classical stellar atmospheric theory applied to a plane parallel geometry compatible with a disk structure.
We combine this procedure with the calculations of the scale-height in the vicinity of the planet (Eq.~\ref{Hplanet}).

The total pressure is assumed to be the contribution of the gas, and radiation pressure, i.e., $P = P_{\rm g} + P_{\rm r}$, respectively. The gas pressure can be written as:
\begin{equation}
    P_{\rm g} = \rho c_{\rm s}^2.
\end{equation}
It is important to note that the isothermal sound-speed $c_{\rm s}$ is related to the gas pressure $P_{\rm g}$ only, and not to the total pressure $P$. The radiation pressure $P_{\rm r}$, can be written as:
\begin{equation}
    P_{\rm r} = \frac{4 \pi}{c} K_\nu,
\end{equation}
where $c$ is the speed of light, and $K_\nu$ is the second momentum radiation flux (proportional to the radiation pressure tensor) and defined as:
\begin{equation}
    K_\nu \equiv \frac{1}{4 \pi} \int I_\nu \mu^2 {\rm d}\Omega,
\end{equation}
where $I_\nu$ is the radiation intensity field in the solid angle $\Omega$ and its direction specified by $\mu \equiv \cos(\theta)$ (being $\theta$ the elevation angle respect to the surface normal). For a semi-isotropic radiation field we have $\mu \simeq 1$\footnote{For a detailed discussion on Stellar atmospheres, and the radiative transfer equations consult \cite{Mihalas1978_book, Mihalas+1984}.}.

Assuming an energy balance in the vertical direction in which the energy dissipated per unit volume $Q^+$ equals the net radiation loss per unit volume $Q^-$, from the total pressure $P$ of the fluid, and  the definition of the moment equation $K_\nu$ from the stellar atmospheres theory \citep{Mihalas1978_book}, the hydrostatic equilibrium condition in Equation \ref{hidroeq} can be rewritten as a density equation \citep{Hubeny1990}:
\begin{equation}\label{d_rho_dz}
    \frac{{\rm d}\rho}{{\rm d}z} = - \frac{2}{H_{\rm g}^2} \rho z + \frac{2 H_{\rm r}}{H_{\rm g}^2} \rho [  1 - \theta(z)],
\end{equation}
where $H_{\rm g}$ is defined as the pure gas pressure scale-height, given by:
\begin{equation}\label{Hg}
    H_{\rm g} \equiv \begin{cases} H_{\rm planet} \mbox{~ (Eq. \ref{Hplanet})}, & \mbox{if } s<s_t \\ H_{\rm g} \mbox{~ (Eq. \ref{Hstd})}, & \mbox{if } s \geq s_t, \end{cases}
\end{equation}
and $H_{\rm r}$ represents a pure radiative pressure scale-height, given by:
\begin{equation}\label{Hr}
  H_{\rm r} \equiv (\sigma/c) T_{\rm eff}^4 \kappa/\Omega_{\rm K}^2,  
\end{equation}
where $T_{\rm eff}$ is the disk effective temperature, $\Omega_{\rm K}$ the Keplerian velocity, $\kappa$ the flux mean opacity, and $\sigma$ the Stefan-Boltzmann constant. The effective temperature is related with the mid-plane temperature through:
\begin{equation}\label{EqTempEff}
 T_{\rm eff}^4=\frac{T^4}{\tau_{\rm eff}},
\end{equation}
where we use for the effective optical depth $\tau_{\rm eff}$ (valid for optically thick case, \citealp{Hubeny1990}): 

\begin{equation}\label{EqTauEff}
 \tau_{\rm eff}= \frac{\sqrt{3}}{4} + \frac{3\tau}{8} + \frac{1}{4\tau},
\end{equation}
and the optical depth is obtained from the mean opacity $\kappa$:
\begin{equation}\label{tau_vol}
    \tau = \frac{1}{2} \kappa \rho.
\end{equation}

To solve Equation \ref{d_rho_dz} one can use a monotonically increasing function between 0 and 1 for $\theta(z)$:
\begin{equation}\label{thetaz}
\theta(z) = \begin{cases} 1 - (z/H), & \mbox{if } \mbox{$z < H$} \\ 0, & \mbox{if } \mbox{$z \geq H$,} \end{cases}
\end{equation}
where $H$ may be called the density scale-height, given by:
\begin{equation}\label{H}
    H = \Sigma / (\sqrt{2 \pi} \rho_0),
\end{equation}
where $\Sigma = \int_0^\infty \rho(z) {\rm d}z$ corresponds to the integrated surface density, and $\rho_0 = \rho(z=0)$.

Using Eqs.~\ref{H} and \ref{thetaz}, the differential density Eq.~\ref{d_rho_dz} can be solved to obtain $\rho(z)$:

\begin{equation}\label{rho_z}
 \rho(z) = \begin{cases} 
\rho_0 \exp{ \left\{ - \left[ \frac{1}{2}  \frac{-z^2}{H_{\rm g}^2} \left( 1 - \frac{H_{\rm r}}{H} \right) + \frac{|z|}{H_{\rm palanet}} \right] \right\}}, & \mbox{if } \mbox{$z < H$} \\ 
 \rho_0 \exp{ \left\{ -  \left[  \frac{1}{2} \left( \frac{z - H_{\rm r}}{H_{\rm g}} \right)^2 + \frac{|z|}{H_{\rm planet}} \right]  \right\}}  \exp{ \left\{ -\frac{(H - H_{\rm r})}{H_{\rm g}} \frac{H_{\rm r}}{H_{\rm g}} \right\}  }  ,         & \mbox{if } \mbox{$z \geq H$.} 
 \end{cases}
\end{equation}

The term $|z|/H_{\rm planet}$ in the above equation is only considered in regions where the planet gravity dominates, i.e., when $s<s_t$.

We have now all the ingredients to derive the density scale-height $H$ of the disk. We follow the treatment from \cite{Hubeny1990} by introducing the dimensionless parameters:
\begin{equation}\label{dimensionless}
    h \equiv H / H_{\rm g}, \, \, \, \, h_{\rm r} \equiv H_{\rm r} / H_{\rm g},
\end{equation}
which allows us to construct an algebraic equation to compute $H$ in the form of the dimensionless parameter $h$:
\begin{equation}\label{h}
\begin{array}{lcl}
    h  &=& \frac{\sqrt{\pi}}{2} \left( \frac{h}{h-h_{\rm r}} \right)^{1/2} [1 - {\rm erf}(\{ h(h-h_{\rm r})\}^{1/2})] \\
    & & + \, {\rm erf}(h-h_{\rm r}) \exp{[-(h-h_{\rm r})h_{\rm r}]},
\end{array}
\end{equation}
where ${\rm erf}(x) = \int_x^\infty \exp{(-t^2)} {\rm d}t$ is the error function. 

Once a numerical solution for $h$ is found, the final scale-height is simply given by:
\begin{equation}\label{Hrad}
    H \equiv H_{\rm g+r} = h H_{\rm g},
\end{equation}
where $H_{\rm g}$ is defined in Equations \ref{Hg}. We call $H_{\rm g+r}$ the density scale-height obtained from Equation \ref{Hrad} and \ref{h}, which corresponds to the scale-height, containing contributions from both gas and radiation pressure. 
It is worth to mention that $H_{\rm r}$ in Equation \ref{Hr} represents the semi-thickness of the disk only when radiative forces are dominant. When the radiation pressure is negligible, i.e. $H_{\rm r} \ll H_{\rm g}$, the scale-height is given by:
\begin{equation}
    H \simeq  H_{\rm g},
\end{equation}
recovering the standard calculation defined in Eq. \ref{Hg}

In this limit, the vertical density profile in Equation \ref{rho_z} has the standard form:

\begin{equation}
    \rho(z) \simeq \begin{cases} \rho_0 \exp\{- \frac{1}{2} (z/H_g)^2\}, & \mbox{if } \mbox{$s \geq s_t$} \\ \rho_0 \exp\{-[ \frac{1}{2} (z/H_g)^2 + |z|/H_{\rm planet} ]\}, & \mbox{if } \mbox{$s < s_t$,} \end{cases}
\end{equation}
where $\rho_0 = \Sigma / (\sqrt{2 \pi}  H_g)$. 

Summarizing, $H_{\rm g+r}$ (Eq. \ref{Hrad}) represents the circumstellar scale-height when gas and radiation pressures are taken into account, and $H_{\rm g}$ (Eq. \ref{Hg}) when only gas pressure is included.

\section{Numerical hydro-simulations}\label{numerical}

\begin{figure*}
\centering
\includegraphics[width=7cm,scale=0.5]{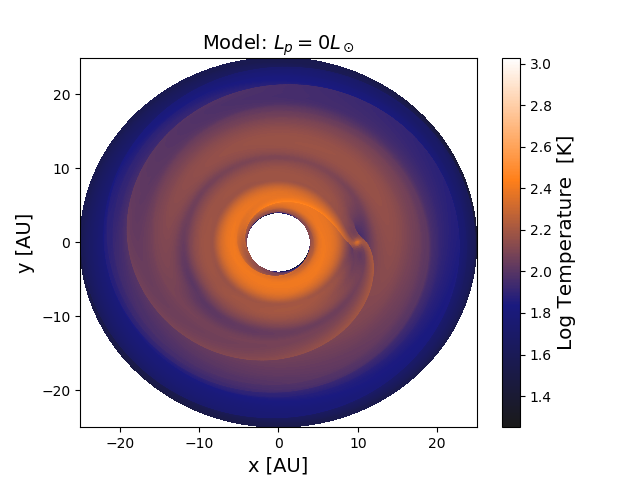}
\includegraphics[width=7cm,scale=0.5]{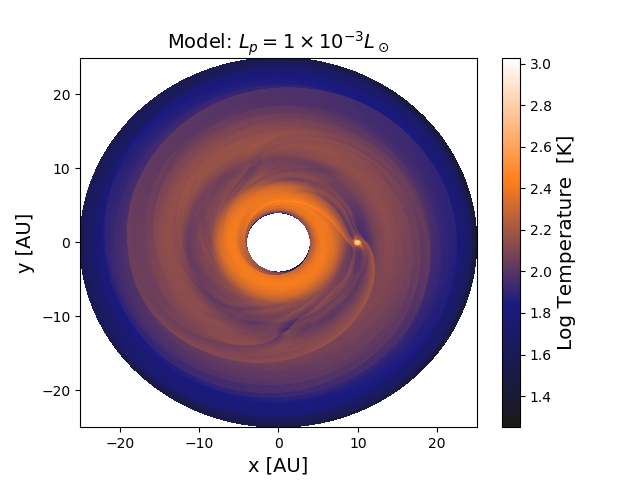}
\caption{Temperature profile of the disk after $10^4$ years. The left panel shows a model without planet feedback $L_p = 0$ (temperature peak $\sim 200$ K), while the right panel includes a $L_{\rm p} = 1 \times 10^{-3} L_\odot$ luminous planet (temperature peak $\sim 1060$ K).}
\label{T2d}
\end{figure*}

\begin{figure}
\centering
\includegraphics[width=7cm]{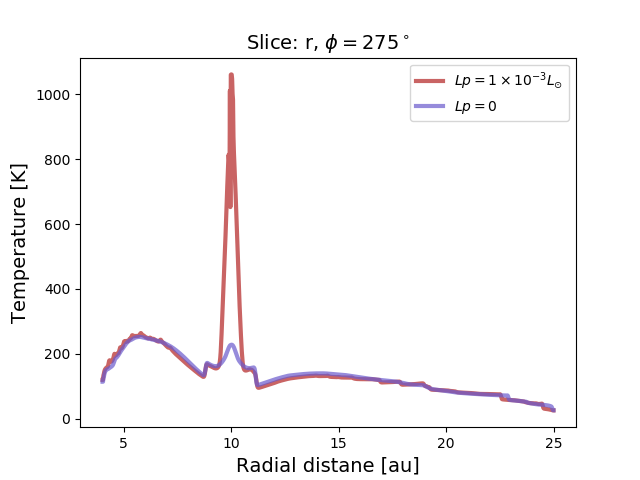}
\caption{Slice of the temperature field along the radial direction --- passing through the planet located at $r_p = 10$ au --- for models with and without feedback in red and blue (respectively).}
\label{temp2d}
\end{figure}

We ran a set of hydro-simulations to follow the evolution of a gaseous disk with an embedded Jupiter mass planet, in which the planet radiates away a fraction of its internal energy at a constant rate.  We modify the the public {\sc FARGO-ADSG} code \citep{Baruteau-Masset-2008} following the same procedure as in \cite{Montesinos+2015, Montesinos+2016}. Assuming hydrostatic equilibrium, we computed the vertical scale-height following the expression for $H_{\rm g+r}$ (Eq. \ref{Hrad}) to recreate a 3D structure, with an initial aspect-ratio $H/r$ set to 0.05 for each model. 
The gravitational force of the planet uses a softening length $\epsilon = 0.6$ over which the potential is smoothed to guarantee a good agreement between 2D and 3D simulations \citep{Masset2002, Baruteau-Masset-2008}.  

We solve a non-stationary energy equation considering a heating term due to shear viscosity, where we adopt the $\alpha$-viscosity prescription \citep{Shakura+Sunyaev1973}, with $\alpha = 10^{-4}$. For the radiative cooling mechanism we assume black-body emission $Q^- = 2 \sigma T^4_{\rm eff}$. We also take into account a radiative heating source (feedback) from the planet as in \cite{Montesinos+2015}.

The disk extends from 4 to 25 au, with an initial surface density given by:
\begin{equation}\label{DensUnper}
 \Sigma(r) = \Sigma_0 \frac{r_\textrm{p}}{r} \,\,\,,
\end{equation} 
where $\Sigma_0 = 30 ~  \rm g ~ cm^{-2}$ corresponds to the density at the planet location $r_{\rm p}$. The total disk mass is therefore $M_{\rm disk} \approx 5 \times 10^{-3} M_\odot$. The embedded luminous planet ($L_{\rm p} = 10^{-3} \, L_\odot$) is located at 10 au. 

The grid resolution for all the simulations was set to $n_r = 1024$ logarithmically spaced radial sectors, and $n_\theta = 1024$ azimuthal equally spaced sectors over $2 \pi$. We initially put a planet located at $r_{\rm p} = 10$ au, $\phi_{\rm p} = 270^\circ$ on a circular orbit ($\phi=0^\circ$ is the north). The planet is not allowed to migrate. We follow the evolution of the system in the reference frame of the planet. We present here two cases: one in which the planet feedback is turned-off (i.e. $L_{\rm p} = 0$), and another one in which the planet radiates at a constant luminosity $L_{\rm p} = 1 \times 10^{-3} \, L_\odot$, which we have shown is achievable even when considering accretion feedback \citep{Garate2020}. 

\begin{figure*}
\centering
\includegraphics[width=5cm]{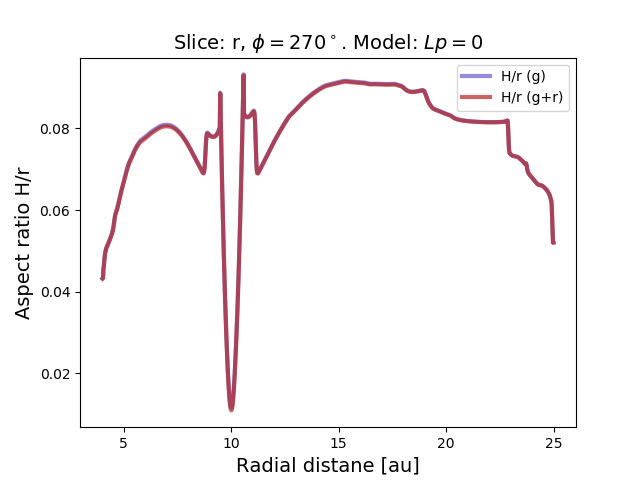}
\includegraphics[width=5cm]{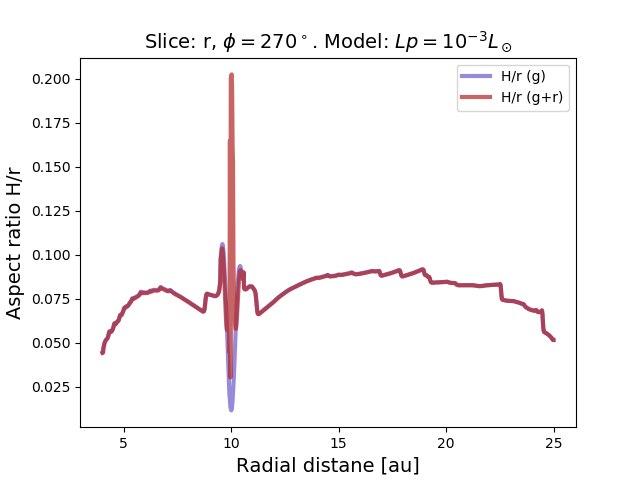}
\includegraphics[width=5cm]{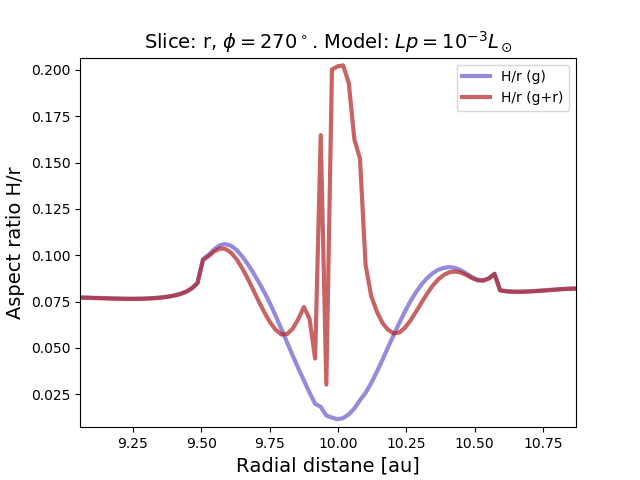}

\caption{Radial slices of the aspect ratio $h=H/r$ --- passing through the planet location --- computed with the pure gas formula $h_{\rm g} = H_{\rm g} / r$ (blue) from equation \ref{Hg}, and the radiative one $h_{\rm g+r} = H_{\rm g+r} / r$ (red) from Equation \ref{Hrad}. The Left panel corresponds to a model without feedback. In this case both curves ($h_{\rm g}$ and $h_{\rm g+r}$) matches. The middle panel corresponds to a model with planet feedback. The radiative component in the expression $h_{\rm g+r}$ overcomes gravitational forces from the planet, locally enhancing the aspect ratio. The right panel is a zoom of the vicinity of the planet (around $\sim 10$ au) for the feedback model. The observed fluctuation around $\sim 9.95$ au responds to a perturbation in the temperature field due to the planetary feedback.}
\label{aspect2d}
\end{figure*}

Figure \ref{T2d} shows the mid-plane temperature field after $10^4$ years of disk evolution (or $\sim$ 316 planetary orbits) at which the system has reached a quasi-steady state. The left panel shows a model without feedback ($L_{\rm p} = 0$). In this case, the gas temperature at the planet location is about $200$~K. When the feedback is included, the temperature peaks at $1060$~K at the circumplanetary regions (right panel).

We are interested in the temperature increment at the planet location. Hence, in Figure \ref{temp2d} we present a slice corresponding to the last simulation orbit of the temperature profile as a function of the radial distance $r$ passing through the planet at fixed  $\phi_{\rm p}$. As in Figure \ref{T2d}, we observe that when the feedback is turned-on, the temperature peaks at 1060~K, while it peaks at 200~K when the planet is not emitting. 

In Fig.~\ref{aspect2d} we show how the disk aspect-ratio $H/r$ is affected by temperature changes due to the planet feedback. We plot $H/r$ for a slice passing over the planet, using the pure gas scale-height $H_{\rm g}$ model (Eq.~\ref{Hg}) and the radiative scale-height $H_{\rm g+r}$ (Eq.~\ref{Hrad}). The left panel of Figure \ref{aspect2d} corresponds to a model with no feedback. In this case, both calculations ($H_{\rm g}$ and $H_{\rm g+r}$) matched, indicating that radiative pressure is negligible when the feedback is turned off. In the right panel of Fig.~\ref{aspect2d}, we plot the model with a luminous planet. In this case, the radiative pressure locally overcomes the planet gravity, enhancing the scale-height above the planet. The CPD region increased its height --- with respect to  $H_{\rm g}$ --- by $(H_{\rm r} - H_{\rm g})/H_{\rm g} \sim (0.2-0.01)/0.01 = 19$. Radiative forces are dominant above the CPD when the planet feedback is activated. 

\section{Observational effects of the radiative scale-height}\label{observations}

\subsection{Vertical optical depth}

\begin{figure}
\centering
\includegraphics[width=9cm]{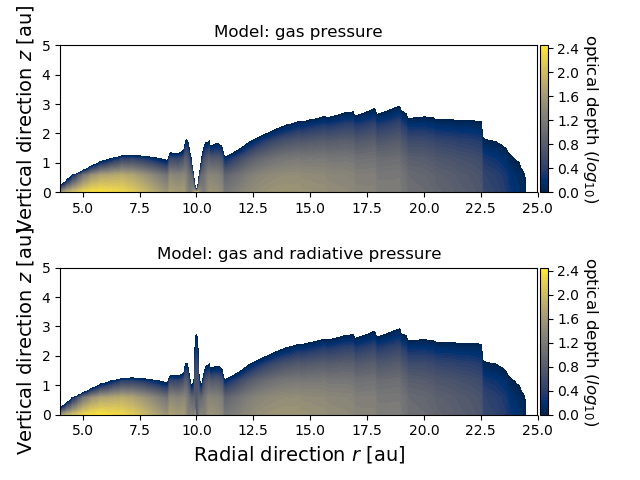}
\caption{Optical depth $\tau(r, z)$ of the disk as a function of the vertical and radial direction for a model with $L_p = 10^{-3} L_\odot$. The top panel was computed using the pure gas formulation $H_{\rm g}$ (Eq. \ref{Hg}), the bottom panel uses $H_{\rm g+r}$ (Eq. \ref{Hrad}). At the planet location ($r_{\rm p} = 10$ au) an optically thick column of gas (a bump) is created due to the gas pressure of the circumplanetary disk. When radiation pressure is included, the \textit{bump} is enhanced by a factor of 19.} 
\label{opticalr}
\end{figure}

To create a 3D distribution, we vertically extend the disk using Equation \ref{Hrad}, where the volume density $\rho(r,\phi, z)$ is computed from Equation \ref{rho_z}. A scale-height enhancement as the one observed in Figure~\ref{aspect2d} (right panel) could act as a bump blocking or scattering a fraction of the stellar radiation. Therefore, it is expected to cast a shadow into the outer disk in the radial direction away from the planet. However, this phenomenon only occurs if, and only if, the ``bump'' is optically thick in the vertical and radial directions.\footnote{in this case ``radial'' refers to a spherical coordinate, rather than the cylindrical $r$ coordinate used in the simulations.}  

We compute the optical depth in Equation \ref{tau_vol} from the volume density given by Eq.~\ref{rho_z}, using for $\kappa$ the Rosseland mean opacity appropriate for protoplanetary disks \citep{Semenov+2003}. Figure~\ref{opticalr} shows a slice cut of the vertical optical depth $\tau(z, r) = \int_z^\infty \frac{1}{2} \kappa(r) \rho(r,z) dz$, passing through the planet for the model with planetary feedback. We note that, when using the radiative formulation $H_{\rm g+r}$ (Eq.~\ref{Hrad}), an optically thick ``bump'' of about $\sim 2$ au of height forms above the CPD (the planet is at $r_p = 10$ au), compared with a \textit{no-bump} situation if the pure gas formulation $H_{\rm g}$ (Eq.~\ref{Hg}) is used, meaning that planet gravity dominates over gas pressure alone.

As mentioned before, if the ``bump'' is also optically thick in the radial direction, then a shadow is expected to be cast. We explore this possibility in the next section through radiative transfer calculations.

\begin{figure}
\centering
\includegraphics[width=9cm]{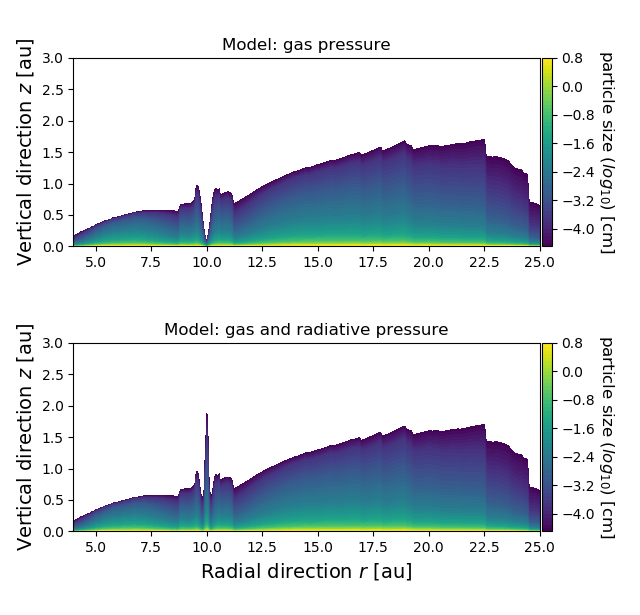}
\caption{Grain size ($10^{-4} - 10$ cm) distribution in $r, z$ for a model with $L_p = 10^{-3} L_\odot$. The slice passes through the planet located at $r_p = 10$ au. The top figure shows a model computed with $H_{\rm g}$, the bottom one includes radiative pressure $H_{\rm g+r}$. The upper layers of the disk are populated with micron-size particles as expected, while biggest particles ($\sim$ cm) settle to the mid-plane.}
\label{sizes}
\end{figure}

\subsection{Monte Carlo radiative transfer: Dust vertical structure}

\begin{figure*}
\centering
\includegraphics{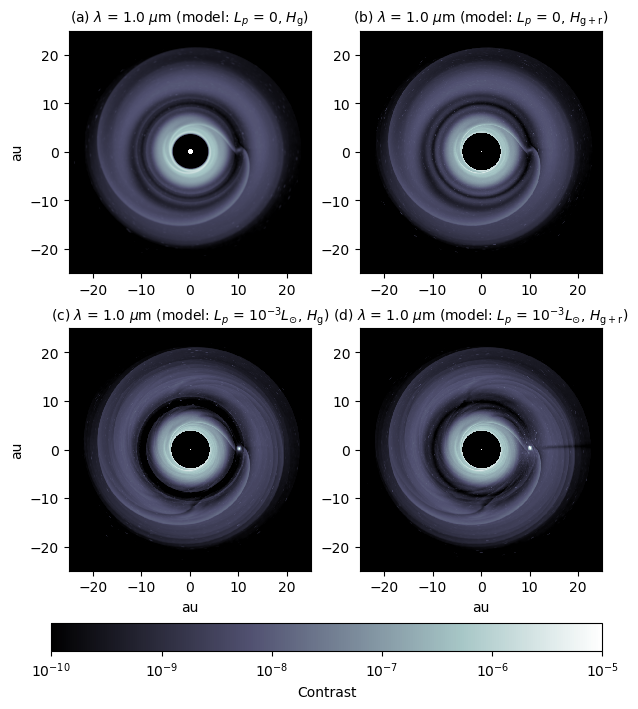}
\caption{Images for $\lambda = 1 \, \mu$m obtained from {\sc radmc3d}. Panel a): model without feedback ($L_{\rm p} = 0$), using the pure gas model $H_{\rm g}$. Panel b): model with $L_{\rm p} = 0$, but using the new formula $H_{\rm g+r}$. Panel c): model with feedback $L_{\rm p} = 1 \times 10^{-3} \, L_\sun$, using the pure gas formula $H_{\rm g}$. Panel d): same feedback as c), but using $H_{\rm g+r}$. A shadow cast from the planet location is observed in this model.}
\label{1micron}
\end{figure*}

We perform radiative transfer calculations using the radiative transfer code {\sc radmc3d} \citep{radmc3d}. We feed {\sc radmc3d} with the gas distribution obtained from our hydro-simulations (see section \S{\ref{numerical}}), corresponding to an evolutionary step after $\approx 10^4$ yrs. The dust  of the disk is assumed to be composed by astrosilicates of intrinsic density  $\rho_{\rm intr}=2 \, {\rm g}\,{\rm cm}^{-3}$ following a power-law size distribution $ {\rm d}n(a) \propto a^{-3.5} {\rm d}a$, with $a$ ranging from 0.1 $\mu m$ to 1 cm. The dust density is normalized by imposing a gas-to-dust ratio of 100. We assume local thermodynamic equilibrium, i.e. $\rm T_{\rm gas} = T_{\rm dust}$. The absorption efficiencies were computed using Mie theory \citep{Bohren+Huffman1983}. 

We decompose the dust size range into 12 logarithmically spaced bins representing 12 dust $i$-species. We obtain a dust density distribution $\Sigma_i$ for each $i$-bin in such a way that the sum of individual species surface density gives the total dust density i.e., $\Sigma_{\rm dust}(r,\phi) = \displaystyle\sum_{i} \Sigma_i(r,\phi)$, where the sum is performed from $a_{\rm min}$ to $a_{\rm max}$. The Stokes number is computed from ${\rm St} = \rho_{\rm intr} a \Omega_{\rm K} / \rho c_{\rm s}$, where $\rho_{\rm intr}$ is the intrinsic density of particles, $a$ the particle radius, $\Omega_{\rm K}$ the Keplerian velocity, $\rho$ the volume density, and $c_{\rm s}$ the isothermal sound speed.

The dust scale-height is assumed to be $H_{\rm d} =  H_{\rm g}  \sqrt{\alpha / (\alpha + {\rm St}_i)}$\footnote{from this we have that small grains with small Stokes number (${\rm St} \ll \alpha$) have scale-heights $\sim$ gas scale-height.}, where $\alpha$ is the turbulent viscosity, ${\rm St}_i$ the average Stokes number of the $i^{\rm th}$-species, and $H_{\rm g}$ the density scale-height computed from $H_{\rm g}$ (Eq.~\ref{Hg}) or $H_{\rm g+r}$ (Eq.~\ref{Hrad}). This method allows us to take into account different vertical distributions for different dust species $i$ \citep{Youdin+2007}.

In Figure \ref{sizes}, we plot the dust scale-height $H_{\rm d}(r,z)$ for the case with planetary feedback. The top panel corresponds to computations with $H_{\rm g}$ (Eq. \ref{Hg}), while the bottom one used the generalized $H_{\rm g+r}$ expression (Eq. \ref{Hrad}). When using $H_{\rm g}$ alone, the bump over the CPD disappears.

For the 3D radiative transfer calculation, we assume that the central star is of solar-type with an effective temperature of 6000K. We incline the disk by $13^\circ$, and we set the system at 140~pc from the Earth.

In Figure \ref{1micron} we show the relative intensity ($I/I_{\rm max}$) for $\lambda = 1 \, \mu$m. We compare the effect of the feedback (switched on/off) when using the different scale-height formulations ($H_{\rm g}$ and $H_{\rm g+r}$) to compute the 3D distribution as an input in {\sc radmc3d}. We remark that:
\begin{enumerate}
    \item A shadow is cast in scattered light from the CPD projected to the outer disk only when the feedback is activated, and the formulation  $H_{\rm g+r}$ is used.
    
    \item No shadows are observed when $L_p = 0$, whether $H_{\rm g}$ or $H_{\rm g+r}$ is used to compute the synthetic image.
    
\end{enumerate}

The features described above can be explained as follows. At the CPD location, the gravity from the planet dominates over gas pressure forces. When the feedback is activated (producing a local temperature increment from $200$ to $1060$~K), radiative forces play a relevant role. They can overcome the gravity from the planet, and the material is puffed-up in the vertical direction (Fig. \ref{aspect2d}, left panel). This material is optically thick, producing a narrow shadow cast from the CPD to the outer disk regions (Fig. \ref{1micron}).

We repeat the same calculation for $\lambda = 10 \, \mu$m. In this case, the shadows disappear since large dust tends to settle, making the bump optically thin at those frequencies. The shadow is observed only in reflected light and not in the dust temperature.
This suggests that the ``optimal'' wavelength for shadow detection should be dominated by scattered light ($\sim 1-3 \mu$m) rather than thermal emission ($> 10 \mu$m). However, to validate these results, proper dust modeling is required with large grain sizes ($\geq 100 \mu$m), where the dust may not be well coupled to the gas as assumed in this work. 

It is worth noting in Figure \ref{1micron} that in models with $L_p = 10^{-3} L_\odot$ (bottom panels) some azimuthal features appear, such as wakes within the gap and the disk. Their origin comes from the planetary feedback, which heats-up the gap, producing noticeable turbulence (see \citealt{Montesinos+2015, Garate2020} for a discussion about the feedback).

\begin{figure*}
\centering
\includegraphics[width=7cm]{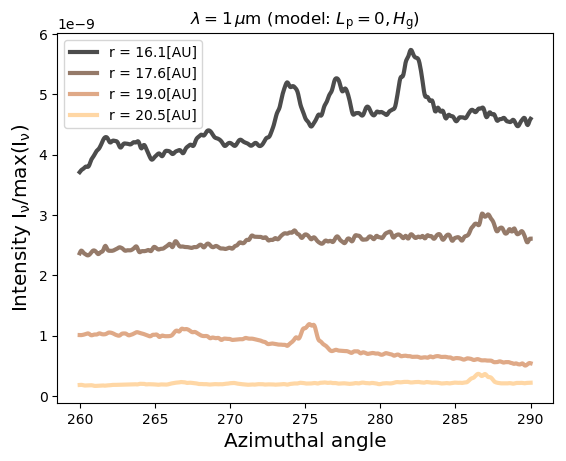}
\includegraphics[width=7cm]{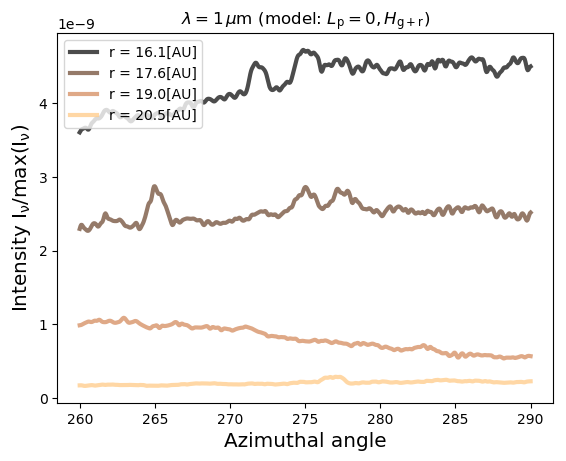}
\includegraphics[width=7cm]{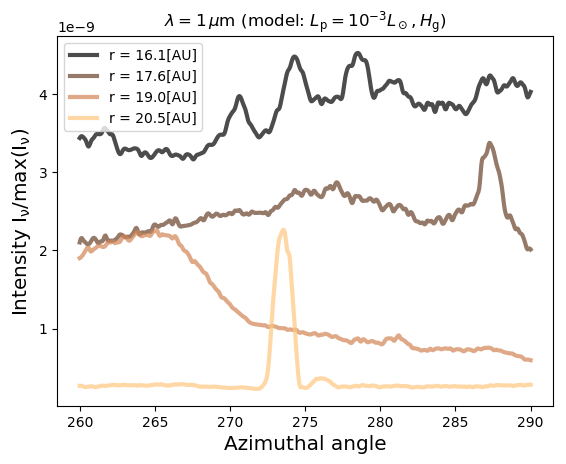}
\includegraphics[width=7cm]{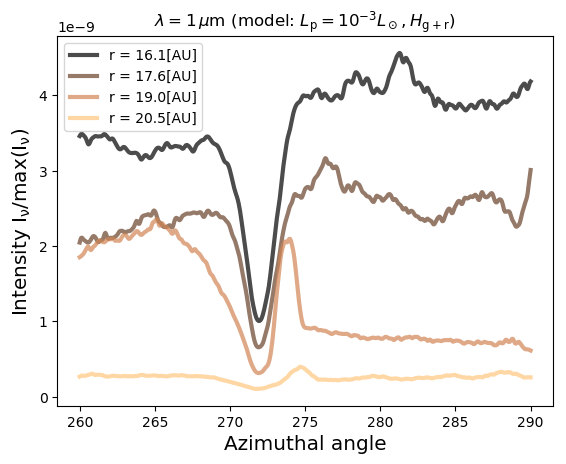}
\caption{Slice cuts between $260^\circ$ to $290^\circ$ degrees for different radial cuts --- ranging from 16.1 to 20.5~au --- of the emissivity reported for $\lambda = 1 \mu$m in Fig.~9. The top panels correspond to models without feedback, where the emissivity was calculated using $H_{\rm g}$ (top left) and $H_{\rm g+r}$ (top right), respectively. The bottom panels include the $10^{-3} L_\odot$ feedback models, with $H_{\rm g}$ (left) and $H_{\rm g+r}$ (right). The feedback model report a strong brightness dip when $H_{\rm g+r}$ is used.
}
\label{intensity_line}
\end{figure*}

In Figure~\ref{intensity_line}, we plot the same disk emissivity reported for $\lambda = 1 \, \mu$m (Figure \ref{1micron}), but following a slice arc between $260^\circ$ to $290^\circ$ for different radii ranging from 16 to 20 au. The top panels of Figure~\ref{intensity_line} correspond to models without feedback, where the emissivity was calculated using $H_{\rm g}$ (top left), and $H_{\rm g+r}$ (top right). The bottom panels include the $L_p = 10^{-3} \, L_\odot$ feedback model, with $H_{\rm g}$ (left) and $H_{\rm g+r}$ (right). Looking at the feedback model with $H_{\rm g+r}$ (bottom, right panel) an emission dip is observed up to 20 au.

\section{Discussion}\label{discussion}

We model the evolution of a protoplanetary disk with an embedded Jupiter mass planet, in which the planet presents an intrinsic luminosity of $10^{-3} L_\odot$.  When this feedback is activated, the region around the planet reaches a mid-plane temperature of about $\sim 1060$ K, such high luminosities are supposed to be present during the early phases of planet formation ($\sim$ Myr old planet) and can last for about $\sim 1$ Myr \citep{Mordasini+2012}.

The disk's vertical structure is described by taking into account the pressure terms related to the gas and the isotropic radiation inside the fluid. The gravity of the planet is also taken into account to counterbalance the pressure gradient from the CPD \citep{Muller+2012}.

The radiation pressure treatment follows an analytic expression adopted from classical stellar atmospheric readjusted to disk geometry \citep{Hubeny1990}. This treatment relies on the assumption of a vertical isothermal sound speed, which is an adequate approximation when the disk interior is optically thick \citep{Armitage2015}.

It is worth mentioning that an equivalent but simplified calculation of the radiative scale-height can also be found in \cite{Montesinos2011}, where the scale-height in that work considers contribution from radiation pressure, disk self-gravity, and turbulent pressure. Results derived from both approximations (\cite{Hubeny1990} and \citealp{Montesinos2011}) applied to the present disk model are equivalent.

We show that in local thermodynamic equilibrium, the radiative pressure that arises in the vicinity of an accreting proto-planet with mid-plane temperatures above 1000 K can overcome the planet's gravity, producing an optically thick bump in its surroundings. For a non-accreting planet ($L_p = 0$), the radiative pressure can be neglected, as both approaches, $H_{\rm g}$ (Eq. \ref{Hg}) and $H_{\rm g+r}$ (Eq. \ref{Hrad}),  produce the same scale-height (see Figure \ref{aspect2d}), and no shadows are observed.

In protoplanetary disks, hot regions can be reached at least in two different locations: at the inner rim of a transitional disk and a planet-forming region. gThe inner rim of a transitional disk is expected to be at temperatures of about $\sim 1500$K \citep{Dullemond&Monnier2010, Vinkovic2014} which could result in a non-negligible contribution of the vertical radiation pressure impacting its scale-height. Here, we do not attempt to describe the inner rim of disks around pre-main sequence stars. However, we suggest the use of $H_{\rm g+r}$ (Eq. \ref{Hrad}) to explain the extremely high aspect-ratios $H/r \sim 0.2$ needed to interpret several non-resolved observations of disks \citep{Natta+2001, Olofsson+2013}. 

When the planet feedback is included, the midplane of the CPD reaches temperatures of about $\sim 1060$K, and the inclusion of radiation pressure in the scale-height equation (Eq.~\ref{Hrad}) dramatically affects the vertical disk geometry. More specifically, by surpassing the planet's gravity, radiation pressure locally enhances its height by 19\% compared to a pure gas pressure model (left vs. right panels in Fig.~\ref{aspect2d}). 

From radiative transfer calculations, we showed that the ``bump'' created above the CPD is optically thick, casting a shadow in scattered light, which extends to the outer regions of the disk in about $\sim 20$ au from the planet. The shadows are only observed at wavelengths of $\sim 1- 3~ \mu$m, mostly because small particles (micron-size) get lifted to the upper disk layers making the bump optically thick at those wavelengths (Fig. \ref{opticalr}). At longer wavelengths, the shadows disappear, as thermal emission tends to dominate over scattered light. For close-in planets, even though they are relatively bright at 1 micron (Fig. \ref{1micron}), their direct detection can be harmed by speckles and other observational artifacts. On the other hand, since shadows extend much further away in the disk, they are more favorable for detection.

To conciliate our findings with observations, recent {\sc zimpol/sphere} images obtained by \cite{Bertrang+2018, Bertrang+2020} revealed small-scale structures could be due to planet-disk interactions. The observations also show a moving narrow surface brightness dip with an azimuthal width of $\sim 20^\circ$. The moving shadow seems to be cast by a large amount of optically thick material that blocks a fraction of the stellar emission, which could be interpreted as an undetected CPD \citep{Bertrang+2018}. Such dips are precisely the kind of predictions obtained with our prescription for the scale-height.

A compelling case is the first direct image of a forming-planet within the gap of PDS 70 \citep{Keppler+2018, Muller+2018}, where an accreting 10 $M_{\rm J}$ mass planet surrounded by a protoplanetary disk was suggested \citep{Keppler+2019}. Also, \cite{Christiaens+2019} recently presented observational evidence of the presence of a CPD around the protoplanet PDS~70~b. Under these circumstances, a shadow should be cast by the putative CPD. In fact, some shadowing has been observed in the outer disk of PDS~70 \citep{Long+2018A}. Its origin remains however unclear. High-contrast observations of micron-sized dust from PDS~70b could, in principle, reveal new shadows arising from the planet-forming region. This system is an exceptional laboratory to test our ideas, where a dedicated model would be needed to predict possible shadowing effects.

Other interesting illumination patterns --- non-related to accreting planets --- have also been reported. For instance, in HD~142527 two diametrically opposed shadows have been discovered in polarized scattered light by \cite{Avenhaus+2014}. These intensity nulls were later explained by the presence of a misaligned inner disk able to block a fraction of the stellar emission \citep{Marino+2015, Casassus+2015, Price+2018}. The main difference with the shadows obtained in our simulations is that the latter cause both a deeper intensity null and a broader azimuthal extension (e.g. $\sim 30^\circ$ for HD~142527 as opposed to $\sim 10^\circ$ in our models). Moreover, it has been shown that these strong shadows can trigger spirals in the gas and dust distribution \citep{Montesinos+2015, Cuello2019}, respectively. 
Also, if the inclined inner disk is precessing, then the projected shadow rotates as well \citep{Facchini+2018}. Remarkably, this could create planetary-like spirals in the gaseous disk at the co-rotating region with the shadow \citep{Montesinos+2018}. Further  observations of intriguing moving shadows have been reported by \cite{Stolker+2017} and \cite{Debes+2017}. However, there are still doubts about their origin. They could be caused by a precessing inner warped disks \citep{Nealon+2019}, or by moving optically thick clouds/bumps at the inner disk rims. In this work, the shadows from the CPD move at Keplerian rotation, and we do not expect any impact on the gas dynamics due to such illumination effects. We emphasize that the obtained dips are not deep  enough to cool down the gas and trigger azimuthal perturbations in the density field, as in \citealp{Montesinos+2016}.

\section{Summary}\label{conslusions}

In this work we showed that for any region in a circumstellar disk with temperatures larger than $\sim 1000$~K, the vertical radiative pressure effect should be taken into account to compute the disk scale-height, which can be approximated by an analytical expression (Eq.~\ref{Hrad}) derived from stellar atmospheres theory adapted to a disk geometry by \cite{Hubeny1990}. 

In the vicinity of an accreting proto-planet, radiative forces surpass the planet's gravitational force, producing an optically thick bump that may cast a narrow shadow into the outer disk. Such shadow should move at Keplerian speed with the planet. For a non-accreting planet, radiative pressure can be neglected, and no shadows are observed.
Due to the narrow width of the predicted shadows ($\sim 10^\circ$), observations at high angular resolution are required to reach high signal-to-noise levels. These dips should be better observed in scattered light at wavelengths close to $\sim 1-3 \mu$m. The shadows are expected to disappear at longer wavelengths since the bump above the CPD becomes optically thin for lower frequencies, and the disk thermal emission contaminates any residual shadowing pattern. 

\acknowledgments
The authors thank the referee for constructive comments and recommendations, which help to improve this paper.
MM acknowledges financial support from the Chinese Academy of Sciences (CAS) through a CAS-CONICYT Postdoctoral Fellowship administered by the CAS South America Center for Astronomy (CASSACA) in Santiago, Chile. NC acknowledges financial support provided by FONDECYT grant 3170680. 
GHMB acknowledges funding from the European Research Council (ERC) under the European Union  Horizon 2020 research and innovation programme  (grant agreement No. 757957). JO acknowledges financial support from Fondecyt (grant 1180395). AB acknowledges support from FONDECYT grant 1190748. CP acknowledges financial support from FONDECYT (grant 3190691). MM, NC, JC, JO, AB and CP acknowledge support from Iniciativa Cient\'ifica Milenio via the N\'ucleo Milenio de Formaci\'on Planetaria. This project has received funding from the European Union's Horizon 2020 research and innovation programme under the Marie Sk\l{}odowska-Curie grant agreement No 210021.



\bibliography{astro}{}
\bibliographystyle{aasjournal}



\end{document}